# Experiments with a LoRaWAN based Remote ID System for Locating Unmanned Aerial Vehicles (UAVs)

Ali Ghubaish[1], Tara Salman[1], and Raj Jain[1]

[1] Computer Science & Engineering, Washington University in St. Louis, St. Louis 63130, USA.

Correspondence should be addressed to Ali Ghubaish; aghubaish@wustl.edu

## Abstract

Federal Aviation Administration (FAA) of the United States is considering Remote ID systems for unmanned aerial vehicles (UAVs). These systems act as license plates used on automobiles, but they transmit information using radio waves. To be useful, the transmissions in such systems need to reach long distances to minimize the number of ground stations to capture these transmissions. LoRaWAN is designed as a cheap long-range technology to be used for long-range communication for the Internet of Things. Several manufacturers make LoRaWAN modules, which are readily available on the market and are, therefore, ideal for the UAVs Remote IDs at a low-cost. In this paper, we present our experiences in using LoRaWAN technology as a communication technology. Our experiments to identify and locate the UAVs systems uncovered several issues of using LoRaWAN in such systems that are documented in this paper. Using several ground stations, we can determine the location of a UAV equipped with a LoRaWAN module that transmits the UAV Remote ID. Hence, it can help identify UAVs that unintentionally, or intentionally, fly into restricted zones.

## I. Introduction

According to the Federal Aviation Administration (FAA), around seven million unmanned aerial vehicles (UAVs) will be sold in the United States by 2020 [1]. UAVs have great potential in many civilian and military applications. Nevertheless, they can hinder public safety and privacy when flying in unauthorized areas. Governments may restrict or forbid UAVs flying in certain areas without prior permissions. Such areas include airports, borders, and many others. In 2016 alone, around 1,800 violations were reported, including UAVs approaching airplanes and disturbing their safety [2-4]. This number has increased by more than one third compared to 2015. Although no catastrophic accident has happened, it is essential to find a solution to reduce these violations.

Many solutions have been proposed for UAVs surveillance such as the mandatory registrations in the FAA registry, geolocation systems, drone guns, signal jammers, sound recognition systems, and visual perception systems. The FAA started a UAV registry in 2015 to locate the owners of UAVs violating any rules [5, 6]. UAV manufacturers use the global positioning system (GPS), which is a satellite-based navigation system owned by the United States, to detect the UAV's location and prevent it from flying in restricted areas [7, 8]. Two drone guns, "Dronegun" and "DroneDefender," have been offered by two different companies to bring down UAVs causing problems [9, 10]. These guns are used to override the signal between the UAV and its remote control, and the UAV is then controlled by the gun controller. However, drone guns require the UAV to be in the line of sight (LoS) of a human with the gun to find the same frequency used by the UAV's remote control to control it. Signal jammers have been used to prevent UAVs from being controlled by their owners when the UAVs enter restricted areas. This forces the UAVs to go back to their configured home point if they lose their control signal. However, jamming affects other wireless devices that use the same frequency band that the UAVs use. This includes 2.4 GHz used by Wi-Fi, which makes this approach inconvenient in most places. UAVs can also be detected by their propeller sound; hence, two different UAV sound recognition systems purposed by Shi et al. and Anwar et al. [11, 12]. The issue with these systems is that it may not efficiently work if an audio jammer device is attached to the UAV. Visual perception systems like Humans' vision, cameras, and proper monitoring may be easier to enforce security in the restricted areas, but these come with cost and maintenance difficulties.

One of the solutions that are being considered by the FAA is to require all the UAVs to have a Remote ID [13, 14]. These IDs will serve as license plates that transmit information to allow authorities to determine the owners of the UAVs and may detect their locations. Remote ID transmission needs a long-haul wireless technology that is cheap enough for low-cost UAVs but still reaches several miles. We believe LoRaWAN is one such technology that can reach from 9 to 18 miles (15 to 30 kilometers) in optimal cases [15, 16]. Hence, deploying a system that uses LoRaWAN protocol can help track the UAVs.

We have developed a prototype and have experimented with LoRaWAN protocol on UAVs. Our goal was to find the feasibility of using this protocol to locate and identify the UAVs. Finding the location of any UAV required us to determine the 3-dimension (3-D) location of the UAV using several ground stations (GSs) listening to the ID broadcasts from the UAV. Upon reception, each GS estimates the distance between itself and the UAV. A minimum of four GSs is required to estimate the location of the UAV in 3-D. A system like this can help law enforcement to be alerted when any UAV flies in a restricted area. We found several issues with using the LoRaWAN protocol in such systems. These issues include the variability of using different LoRaWAN modules, the module's antenna direction, and the battery capacity to run these modules.

The rest of the paper is organized as follows: Section II provides background and related work; Section III discusses system architecture; Section IV shows the experimental implementation and results in detail. The critical issues discovered by our experiments are discussed in Section V. Finally, conclusions and future work are presented in Section VI.

## II. Background and Related Work

This section gives a brief background on the technologies used in the paper. Besides, we discuss some of the earlier related works.

### A. LoRaWAN

A UAV is controlled by a ground-based remote controller via a radio frequency (RF) communication protocol [17]. RF technologies such as LoRaWAN, Zigbee, and 6LoWPAN can be used for communications [18-21]. LoRaWAN is a relatively new technology that is suitable for UAV communications due to its low power, low cost, and long-range reachability. The medium access control (MAC) protocol for LoRaWAN has been standardized by the LoRa Alliance. It uses the LoRa physical layer that enables it to reach long ranges with low power consumption using the chirp spread spectrum modulation [18, 22]. We selected LoRaWAN for location estimation due to its low cost and long-range reachability. Further description of LoRaWAN can be found in [15].

### B. Distance and Location Estimation

Different methods have been explored in the literature for distance estimation. These methods include the time of arrival (ToA), time of flight (ToF), and received signal strength indication (RSSI) [23-26]. ToA method uses elapsed time between sending and receiving a signal between two nodes to measure the distance between them. For instance, GPS uses the ToA between a client node and a satellite to measure the distance between them [8]. The ToF method measures the time for radio signals to bounce back to the GS after being sent to the UAV. This method has been used in aircraft since 1950 [27].

RSSI is a measure of the quality of the signal and can be used for distance estimation. It measures the power level of the received signal [28]. Its value is measured in decibel (dB) and has multiple applications in wireless communication. One of these applications is distance estimation between two nodes, such as the UAV and the GS [29].

Location estimation of any UAV requires knowing its distance from several GSs with known coordinates. For locating the UAV in 2-dimension (2-D), distances from at least 3 GSs are required. For location estimation in 3-D, distances from four GSs are required. Given the coordinates of the required number of GSs and by estimating the distance using one of the previously stated methods, the location of the UAV can be estimated. For example, the ToA method is being used in GPS, which consists of around 31 satellites [8]. Each satellite broadcasts its location and time. By knowing how far the UAV is from one satellite, the UAV knows its distance from that satellite and knows that it is located on a sphere with the estimated distance as a radius. Adding at least two more satellites' information can help the UAV estimates its location in 2-D by finding the points where the three satellites' spheres intersect. Further, adding more satellites' information to the equation can pinpoint the UAV's location and reduce the uncertainty (error) to a few meters.

### C. Related Work

Most prior works in UAV location estimation use GPS. UAVs can be used for many applications such as delivering products and acting as a flying ad-hoc network for broadband wireless access during emergencies [30-32]. Most of these applications need to know the location of the UAV, and they use GPS coordination for that. However, GPS is not always available and not usable for identification. Thus, investigating other alternative localization solutions with an identification feature is desirable for UAV localization in all applications.

Wang et al. investigate a UAV rescue system, named GuideLoc, that helps to rescue people during a natural disaster using UAVs [33]. GuideLoc captures the average RSSI value of a wireless device signal such as a mobile phone carried by a trapped person. The system uses the antennas attached to the UAV to capture the average RSSI value. If the average RSSI value is less than a threshold, the angle of arrival of the signal gets updated to find the location of that person and to record the GPS coordinates of the trapped person. The angle of arrival is determined by the strength of the average RSSI value. Lee et al. utilize the same technique to localize the sensor nodes in the wireless sensor networks [34]. Our system differs from GuideLoc by relying only on the RSSI values to estimate the UAV location and not the GPS.

Raimundo et al. examine the possibility of using the LoRaWAN communication protocol for a UAV location system [35]. The system consists of a UAV that uses a global navigation satellite system (GNSS) receiver to gather the GNSS data, then sends them by a LoRaWAN module attached to the UAV. GNSS receivers can connect to different satellite-based systems such as GPS and other navigation systems [36]. The LoRaWAN module sends the GNSS positions to a base station on the ground. In our system, the UAV is located and identified using the RSSI values and a message that is broadcasted using the LoRaWAN technology.

UAVs have been used by Ferreira et al. to find the network distribution and coverage in remote areas or hazard locations [37]. The proposed system uses the UAV's center-modem to detect the network access points (APs) using the RSSI values broadcasted by the APs in the network. The system uses these RSSI values to estimate the APs locations based on known UAV locations in different reference points, during the UAV flying path, and the estimates distance to these APs. The free-space propagation model is utilized in the system for distance estimation, and three different location methods are tested [38]. They conclude that Bound Box method has the lowest estimation error with a low variance when increasing the number of reference points. Another system by Greco et al. is similar to that by Ferreira et al., but they rely on radio-frequency identification (RFID) tags instead of APs to be located by the UAVs [39].

One of the issues facing location-based systems is to locate objects or UAVs in indoor environments. Tian et al. introduce the HiQuadLoc system that uses Wi-Fi access points to locate a UAV in an indoor environment [40]. Twenty APs are utilized in the system in an area of 1100 $m^2$. The system uses two phases – an offline phase and an online phase. The offline phase divides the indoor area into cubes with known RSSI values to correctly help detect the UAV location in the online phase. The system achieves an average error of 1.64 m. The UAV speed is varied up to three meters per second. They conclude that the location error increases as the UAV speed increases.

Cheng et al. propose a system that can locate a non-line of sight (NLOS) UAV in an indoor environment [41]. The system uses RSSI values in the NLOS identification algorithm to identify the propagation conditions. Also, they use particle swarm optimization-based maximum joint probability algorithm to find the UAV's 2-D coordinates. The system achieves an average error of 0.85 m.

Our system also uses RSSI values for distance estimation; however, we target outdoor environments rather than indoors, and we use LoRaWAN to allow location estimation over much longer distances.

## III.    System Architecture

In this section, the system components, distance estimation modeling, and location estimation for the RSSI method are discussed. The discussion also includes the modeling methods used to estimate the distance from the RSSI values, along with graphs that illustrate that method.

### A. Prototype Components

As shown in Figure 1, our prototype system consists of five main components: LoRaWAN modules, GSs, antennas, a battery, and a UAV. In the following, we briefly discuss these components:

- **LoRaWAN Modules:** Two different modules are used - Moteino LoRa and Seeeduino LoRaWAN modules for our prototype, as shown in Figure 1. Seeeduino module uses 433/868 MHz frequency bands while the Moteino module uses the 915 MHz band. Both modules can report the RSSI values while the Moteino module has an external antenna for longer ranges. The details of these modules can be found in [42, 43]. Alternatively, we could have used Libelium LoRaWAN module [44]. However, we have used Seeeduino and Moteino as shown in the distance estimation modeling since they meet our requirements such as reporting RSSI values where Libelium module lacks this feature.
- **GS:** For each GS, we use a regular computer connected to a LoRaWAN module. The computer is used to program the LoRaWAN module and record the data.
- **Antenna:** Moteino LoRa module requires a separate directional antenna to work, while the Seeeduino LoRaWAN module has a built-in wire antenna on the module.
- **Battery:** Any power bank is sufficient to power the LoRaWAN module connected to the UAV.
- **UAVs:** We use two different UAVs for the prototype: DJI Phantom 2 and DJI Phantom 4 Pro. As discussed earlier, a LoRaWAN module and a battery have been attached to each UAV.

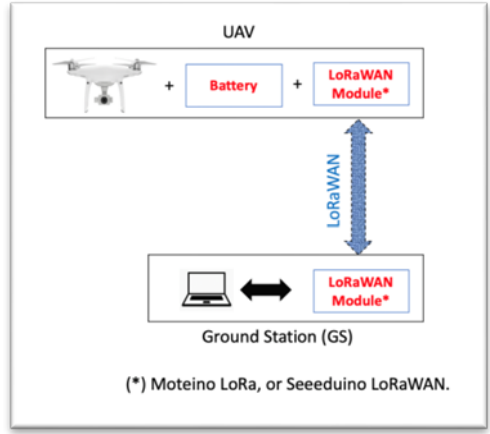

Figure 1: System architecture for distance estimation.
Two LoRaWAN modules, Moteino, and Seeeduino LoRaWAN, are used for distance estimation.

### B. Modeling for Distance Estimation

Using the configuration shown in Figure 1, we estimate the distance between one of the GSs and the UAV using the RSSI values and the log-distance path loss model as will be discussed. For each LoRaWAN module, two modules are used: one is attached to the UAV and powered by a battery, and the other is connected to a computer to control and record the data and serves as the GS.

For distance estimation, the UAV continuously broadcasts a message that has its ID. The interval time of successive messages is two seconds, which is the minimum interval time for the LoRaWAN modules to avoid losing messages [45]. The message length and its effect are explained later in Section IV.

The log-distance path loss model states that [46]:

$$RSSI = -10 * L * log_{10}(d) - C \quad (1)$$

Where *RSSI* is the RSSI values measured at the destination, *d* is the distance, *L* is the path loss exponent, and *C* is a constant. Given (1), the distance between the UAV and the GS can be measured as follows:

$$d = 10^{-(\frac{RSSI-C}{10L})} \quad (2)$$

However, the measured RSSI values can fluctuate, and thus, using one value is not sufficient to estimate the distance. Typically, multiple values need to be used. In our experiments, we used an average of five RSSI values to measure the distance. Five is chosen arbitrarily as a tradeoff between the time and the fluctuation in the RSSI values.

That is, the distance between a GS and the UAV can be estimated as:

$$d = 10^{-(\frac{meanRSSI-C}{10L})} \quad (3)$$

Here, *meanRSSI* is the average RSSI value of five RSSI values.

Even though *C* and *L* are constants in (1), their values are initially unknown and depend on the environment, as discussed by Sherazi et al. [47]. To estimate these parameters, we need *meanRSSI* values and their corresponding distances for a few known positions. Therefore, *a model is needed to estimate these values using (1)*. To do so, we fit a linear model to the *meanRSSI* values. In the resulting linear model, the slope is *-10*L* (thus *L= -slope/10*), and it can be calculated as:

$$slope = \left(\frac{\sum xy - n\bar{x}\bar{y}}{\sum x^2 - nx^{-2}}\right) \quad (4)$$

Where *x* is the *meanRSSI* value at a known position, or a known distance. *y* is the $log_{10}(d)$ value corresponding to the *meanRSSI* value, *n* is the number of RSSI values included in that mean, $\bar{x}$ and $\bar{y}$ are the mean over all *meanRSSI* values and the mean over all $log_{10}(d)$ values, respectively.

In the resulting linear model, the intersection point is -*C* (thus, *C=-Intersection*) which can be calculated from the linear model as:

$$intersection = \bar{y} - slope \times \bar{x} \quad (5)$$

To get the distances between the UAV and the GS, we tried to use a laser meter to measure the distance between the two ends. However, it becomes difficult to do such measurements when the actual distance gets above 200 m. In such a case, the UAV gets smaller and harder to detect by the laser meter. Hence, as shown in Figure 2, we measure the ground distance (GD) between a ground point (GP) and the GS to compute the slant distance (SD) between the GS and the UAV, which equals $d$ in (3). The measurement is relatively accurate, as will be shown in Section IV. The slant distance can be estimated as [48]:

$$SD = \sqrt{GD^2 + H^2 - (2 * GD * H * cos(\beta))} \qquad (6)$$

Here, $H$ is the height of the UAV, which is set to 50 m, $GD$ is the ground distance between the $GP$ under the UAV and the GS, and $\beta$ is the angle between the GS and the UAV. The height is fixed to take the distance as the only variable parameter to simplify the measurements. The $GD$ and its corresponding angle are measured using the laser meter. The $GP$ is selected to be directly below the UAV. Thus, the angle between the UAV and the GP is 90-degrees, and it is measured using the laser meter, which is attached to a tripod. The angle $\alpha$ between the $GP$ and the GS is measured with the laser meter. Note that, $\beta$ can be calculated by subtracting $\alpha$ from 90-degrees:

$$\beta = (90 - \alpha) \qquad (7)$$

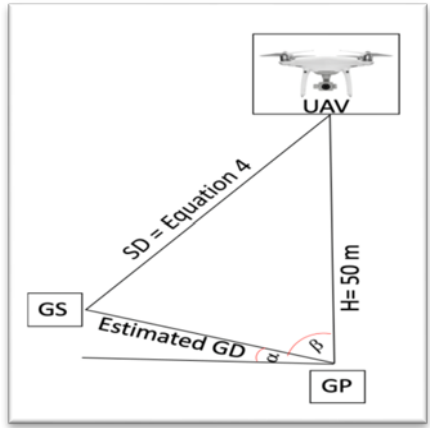

Figure 2: Slant-Distance estimation technique.
This technique is used if the distance between the two nodes is larger than 200 m.

Using the above technique, one can estimate the values for the parameters $C$ and $L$. These parameters can then be used with the measured *meanRSSI* value to estimate distances at other UAV positions.

### C. Location Estimation

In the first stage of our experiments, the Seeeduino LoRaWAN module is used to estimate the location of the UAV using the RSSI method. The location system consists of four GSs and one UAV, as illustrated in Figure 3.

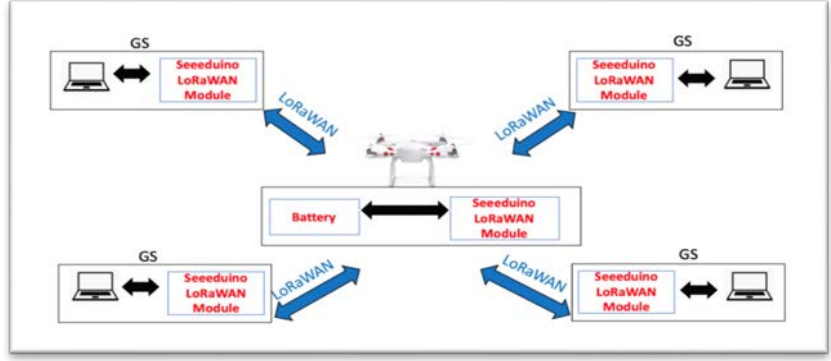

Figure 3: System architecture for the location estimation.

Seeeduino LoRaWAN module is attached to each of the GSs and the UAV. In addition, a battery to power the LoRaWAN module is also attached to the UAV. In each GS, there is a computer that records the *meanRSSI* values received from the module connected to it. In this stage, the data are manually collected from all the four GSs computers and transferred to a fifth computer called the central computer, which is not shown in Figure 3. The transferred data are processed based on three elements: the GSs' 3-D locations, the distances between each of the four GSs, which is 200 m, and the *meanRSSI* values received from the four GSs.

The location estimation uses the SD between the UAV and four GSs. As explained earlier, the Seeeduino module requires an interval time of two seconds between successive messages. To satisfy this requirement and that we need to use the mean of 5 RSSI values, the UAV must stay in one spot for at least 10 seconds. Trilateration technique is used to determine the location of the UAV [49, 50]. This technique has been used to estimate the location in [51-53]. It allows us to determine the exact 3-D location of any object using its distance from at least four points with their known 3-D locations. In our case, the UAV is the object whose location and height need to be determined, while the four GSs are the points with known locations, as shown in Figure 4.

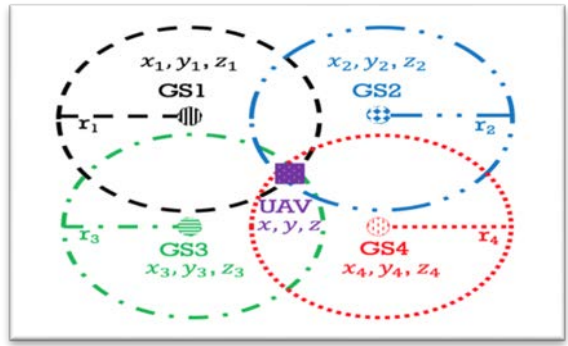

Figure 4: Trilateration system architecture.
Minimum of four GSs 3-D locations is required to find the UAV 3-D location.

The UAV is on the surface of a sphere with radius ($r_i$) centered at $GS_i$. The $r_i$ is equal to $SD_i$ for each GS. The location of the UAV is a 3-element vector *w = {x, y, z}*. It can be computed as the intersection of the four spheres. Each sphere consists of the 3-D location of each GS and the radius value between itself and the UAV. The radius value represents the estimated distance, *SD*, from the previous subsection. Therefore:

$$r_1^2 = (x - x_1)^2 + (y - y_1)^2 + (z - z_1)^2$$
$$r_2^2 = (x - x_2)^2 + (y - y_2)^2 + (z - z_2)^2 \quad (8)$$
$$r_3^2 = (x - x_3)^2 + (y - y_3)^2 + (z - z_3)^2$$
$$r_4^2 = (x - x_4)^2 + (y - y_4)^2 + (z - z_4)^2$$

We can expand out the squares in each one, as shown in (9).

$$r_1^2 = x^2 - 2x_1x + x_1^2 + y^2 - 2y_1y + y_1^2 + z^2 - 2z_1z + z_1^2$$
$$r_2^2 = x^2 - 2x_2x + x_2^2 + y^2 - 2y_2y + y_2^2 + z^2 - 2z_2z + z_2^2 \quad (9)$$
$$r_3^2 = x^2 - 2x_3x + x_3^2 + y^2 - 2y_3y + y_3^2 + z^2 - 2z_3z + z_3^2$$
$$r_4^2 = x^2 - 2x_4x + x_4^2 + y^2 - 2y_4y + y_4^2 + z^2 - 2z_4z + z_4^2$$

By subtracting the 4[th] equation ($r_4$) from the first three equations in (9), we get the following:

$$2(x_4 - x_1)x + 2(y_4 - y_1)y + 2(z_4 - z_1)z = r_1^2 - r_4^2 - x_1^2 - y_1^2 - z_1^2 + x_4^2 + y_4^2 + z_4^2$$
$$2(x_4 - x_2)x + 2(y_4 - y_2)y + 2(z_4 - z_2)z = r_2^2 - r_4^2 - x_2^2 - y_2^2 - z_2^2 + x_4^2 + y_4^2 + z_4^2 \quad (10)$$
$$2(x_4 - x_3)x + 2(y_4 - y_3)y + 2(z_4 - z_3)z = r_3^2 - r_4^2 - x_3^2 - y_3^2 - z_3^2 + x_4^2 + y_4^2 + z_4^2$$

Putting (10) in a matrix form, we get (11) where $A$ is the coefficient matrix, $w$ is a vector of variables to be estimated, i.e., ($x, y, z$) in (10), and $b$ is the right-side vector.

$$\begin{bmatrix} 2(x_4 - x_1) & 2(y_4 - y_1) & 2(z_4 - z_1) \\ 2(x_4 - x_2) & 2(y_4 - y_2) & 2(z_4 - z_2) \\ 2(x_4 - x_3) & 2(y_4 - y_3) & 2(z_4 - z_3) \end{bmatrix} \begin{bmatrix} x \\ y \\ z \end{bmatrix} = \begin{bmatrix} r_1^2 - r_4^2 - x_1^2 - y_1^2 - z_1^2 + x_4^2 + y_4^2 + z_4^2 \\ r_2^2 - r_4^2 - x_2^2 - y_2^2 - z_2^2 + x_4^2 + y_4^2 + z_4^2 \\ r_3^2 - r_4^2 - x_3^2 - y_3^2 - z_3^2 + x_4^2 + y_4^2 + z_4^2 \end{bmatrix} \quad (11)$$
$$A \qquad\qquad w \ = \qquad\qquad b$$

Note that $w$ is the UAV 3-D location that we need to determine given other values in (11). To find $w$, the closed-form of the least-squares method can be used to solve the equation in one step, as shown in (12).

$$w = (A^T A)^{-1} A^T b \quad (12)$$

If the height for all the GSs is the same, the last column of matrix $A$ will be all zeros, and the matrix becomes not invertible. This step can be taken care of by removing the last column of matrix $A$, computing only $x$ and $y$ values from the above equations, and separately determining $z$ as in (13) by substituting $z_4$ value in (8) with zero and solving for $z$. Here, $z$ represents the height of the UAV, while $x$ and $y$ represent the 2-D location of the UAV.

$$z = \sqrt{r_4^2 - (x - x_4)^2 - (y - y_4)^2} \quad (13)$$

## IV. Experimental Implementation and Results

In this section, the experimental implementation and results are discussed. We present the steps to prepare the software and hardware for the experiments. Also, we show some statistical results for the location estimation method.

### A. Distance Estimation Using RSSI Method

Two different outdoor environments were used to model and validate our experiments. All the nodes in the experiment used the Seeeduino LoRaWAN module, which is based on Arduino Zero bootloader with LoRaWAN protocol embedded in it; thus, no additional module was needed [43]. Seeeduino provides a library and examples to use their module. Using these examples, we found that it is possible to vary the contents and formats of the transmitted messages. Hence, three messages with different lengths and formats were tested, as shown in Table 1.

Table 1: Different message lengths and formats list

| #  | Message                                                          | # of Bytes | Format      |
|----|------------------------------------------------------------------|------------|-------------|
| M1 | FF 31                                                            | 2          | Hexadecimal |
| M2 | FF1                                                              | 3          | String      |
| M3 | FF1 is the UAV ID number that is being used to identify this UAV | 66         | String      |

Initially, our test was based on using two nodes mounted on two tripods and not attached to the UAV with distances ranging from 100 to 500 m. As shown in Figure 5, we fitted a linear model to show the relationship between a set of *meanRSSI* values and their corresponding distances using different message lengths. Also, we calculated the confidence interval for each distance and message length to see if different *meanRSSI* values for different distances using one message overlap or not. From Figure 5, we can see that as the message length gets larger, the *meanRSSI* value increases. This finding is essential since, with lower *meanRSSI* values, different distances using different *meanRSSI* values overlap, causing a significant error in distance estimation. For example, by using message 2 (M2), we may get a *meanRSSI* value that could be 100 to 300 m away, resulting in an error of 200 m. Based on this realization, the most extended message among the three messages, M3, was used with the Seeeduino LoRaWAN module to complete the rest of this experiment.

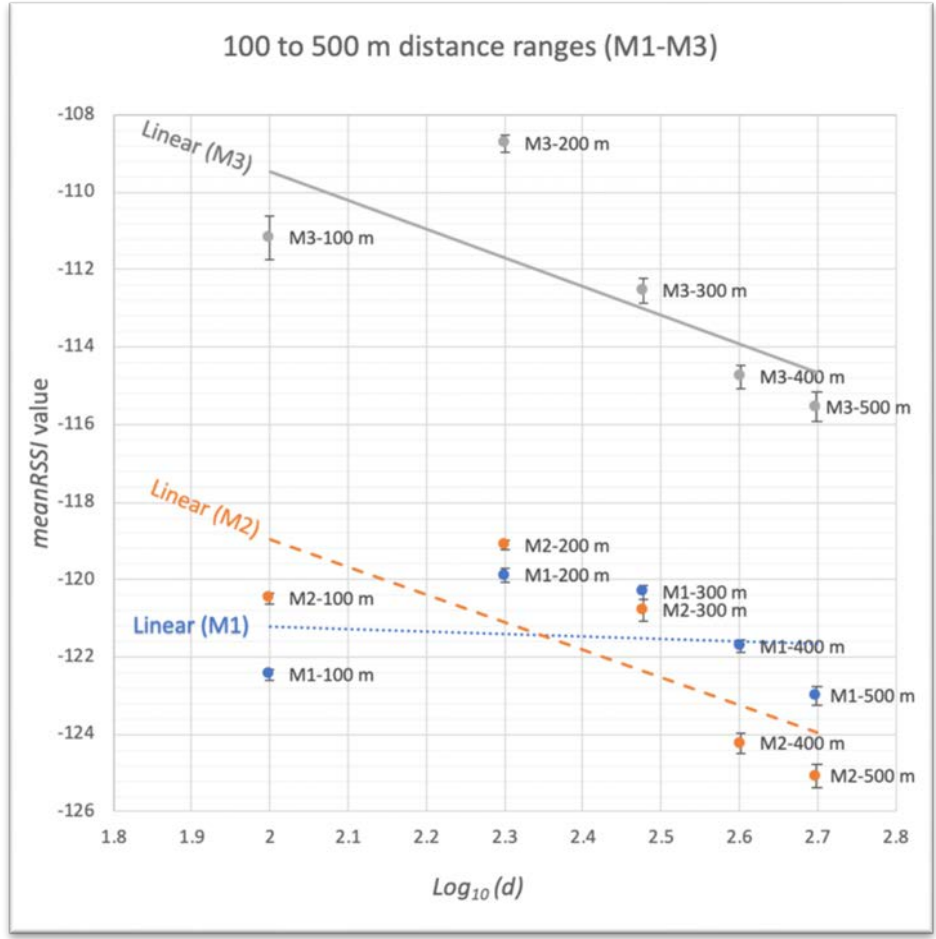

Figure 5: *meanRSSI* values for different message lengths at different distances.

For the second stage, we calculated the SD distances for six different positions with nominal distance ranging from 100 to 600 m, as shown in Table 2. The SD values were those computed using (6). Notice that the calculated SDs were close to the nominal distances.

Table 2: Calculated SDs

| Nominal Distance | Ground Distance (GD) | Height (H) | UAV-GS angle ($\beta$) | GP-GS angle ($\alpha$) | Slant Distance (SD) |
|---|---|---|---|---|---|
| 100 m | 100.0 m | 50 m | 79.1° | 10.9° | 102.97 m |
| 200 m | 200.2 m | 50 m | 81.7° | 8.3° | 199.19 m |
| 300 m | 299.8 m | 50 m | 83.3° | 6.7° | 298.19 m |
| 400 m | 400.3 m | 50 m | 84.0° | 6.0° | 398.15 m |
| 500 m | 500.5 m | 50 m | 84.7° | 5.3° | 498.34 m |
| 600 m | 600.7 m | 50 m | 84.9° | 5.1° | 598.29 m |

After getting the SDs, we performed a statistical analysis on the collected data, as shown in Table 3 using the methods described in [54]. The measurements consist of six *meanRSSI* values; each of which consists of 125 samples in each of the six distance ranges. Initially, the height for the UAV was fixed to 50 m to keep the analysis simple. Then, we conducted another experiment to check if the *meanRSSI* values were the same for different heights up to 100 m. Then, we decreased the GD, and correspondingly we increased the UAV height to keep the same SD. Results showed that the *meanRSSI* values were the same as long the SDs were the same; that is, the *meanRSSI* values were not affected by the height.

Table 3: Statistical characteristics of *meanRSSI* values using the Seeeduino LoRaWAN module

| Nominal Distance | 100 m | 200 m | 300 m | 400 m | 500 m | 600 m |
|---|---|---|---|---|---|---|
| Sample variance | 4.78 | 2.34 | 2.62 | 1.46 | 1.30 | 1.22 |
| Sample Standard Deviation | 2.19 | 1.53 | 1.62 | 1.21 | 1.14 | 1.10 |
| Sample Standard Error | 0.20 | 0.14 | 0.14 | 0.11 | 0.10 | 0.10 |
| Sample Mean ($\bar{x}$) | -79.41 | -82.94 | -85.81 | -85.58 | -87.93 | -88.32 |
| 95% Confidence Interval | (-79.79, -79.03) | (-83.21, -82.68) | (-86.09, -85.52) | (-85.79, -85.37) | (-88.13, -87.73) | (-88.52, -88.13) |
| L | \multicolumn{6}{c}{1.165} |
| C | \multicolumn{6}{c}{-56.134} |
| $R^2$ | \multicolumn{6}{c}{0.97} |

As shown in Table 3, the sample variance decreased as the distance increased. To find the two unknown parameters, *L* and *C*, we used the linear regression model discussed earlier in Section III.B. The results are shown in Figure 6. Note the decreasing variance (and hence narrower confidence interval) as the distance between the UAV and the GS increases. Overall, the model resulted in an $R^2$ value of 97%, which showed that the linear regression model was a good fit.

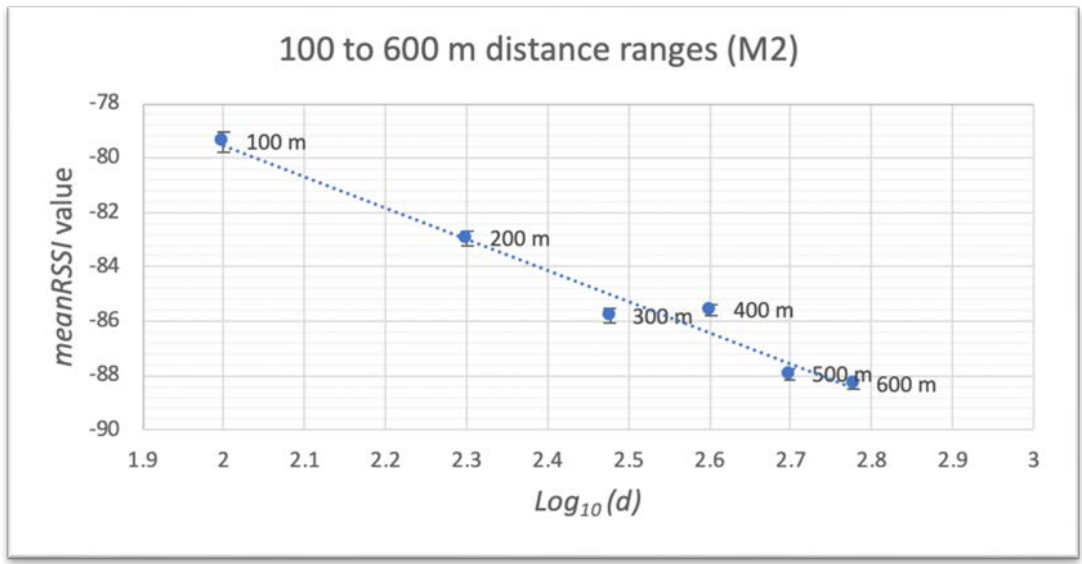

Figure 6: Linear regression model. These measurements are based on the use of the Seeeduino LoRaWAN module.

The confidence interval is essential to see if any *meanRSSI* value for any distance was overlapping with another *meanRSSI* value. After calculating these confidence intervals, we found that their values for 300/400 m values overlap. This overlap showed that the *meanRSSI* values for these two distances were not statistically different. In other words, given a *meanRSSI* value and the calculated *L* and *C*, we may estimate the distance with an error of 100 m, which is a drawback. At this point, we decided to check the *meanRSSI* value with the other LoRaWAN modules (i.e., Moteino LoRaWAN).

As shown in Table 4, the Moteino module produced *meanRSSI* values for different distances that overlap with other distance ranges from 100 to 800 m; hence, it is not a perfect candidate for the linear regression model to find *L* and *C* parameters. We found that the perfect length of the message for the Moteino module was M2 (shown earlier in Table 1). Longer messages, e.g., M3, were transmitted in several fragments. Hence, we ended up using M2 instead of M3.

Table 4: Statistical characteristics of *meanRSSI* values using the Moteino LoRaWAN module

| Distance | 100 m | 200 m | 300 m | 400 m | 500 m | 600 m | 700 m | 800 m |
|---|---|---|---|---|---|---|---|---|
| Sample Mean | -104.48 | -103.68 | -103.51 | -104.97 | -105.06 | -104.86 | -104.62 | -104.65 |
| Confidence Interval | (-104.61, -104.36) | (-103.81, -103.55) | (-103.61, -103.40) | (-105.03, -104.90) | (-105.11, -105.00) | (-104.93, -104.80) | (-104.72, -104.52) | (-104.75, -104.56) |

### B. Location Estimation Using RSSI Method

Location estimation stage consisted of four GSs and one UAV. The UAV used in this stage was DJI Phantom 4. The GSs were 200 m away from each other where all antennas' directions were pointing up since that impacts the *meanRSSI* values, according to Wadhwa et al. [55]. The UAV antenna had a spring shape facing down, as shown in Figure 7. The battery that was used to power-up the LoRaWAN module attached on the UAV was under the module itself as shown in the figure.

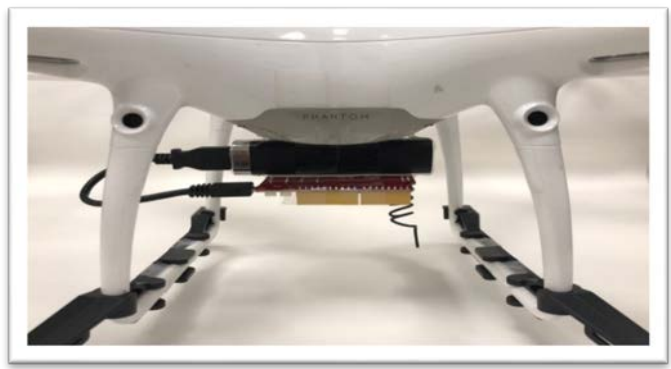

Figure 7: Seeeduino LoRaWAN module with the battery attached to the UAV.
The shape of the antenna and the battery location is essential since the *meanRSSI* values are affected by them.

The measured and estimated SD with UAV at the height of 50 m are shown in Table 5. The real SDs were measured using the GPS, while the estimated SDs were based on the *meanRSSI* values.

Table 5: Statistical characteristics of location determination using the Seeeduino LoRaWAN module

|  | GS1 | GS2 | GS3 | GS4 |
|---|---|---|---|---|
| *MeanRSSI* (dB) | -80 | -86 | -79 | -81 |
| Est. SD (m) | 112 | 366 | 92 | 136 |
| Real SD[1] (m) | 146 | 161 | 140 | 155 |
| Distance Error[2] | 23% | 127% | 34% | 12% |

1   Using Google Maps "measure distance" feature.
2   The error is calculated based on the difference between the estimated SDs and Real SDs.

After we calculated the distance error based on the difference between the estimated SDs and real SDs, we found that GS2 showed a distance error of 127%. This error is discussed further in the next section.

## V.   Issues and Challenges

Although LoRaWAN can be used for distance estimation using RSSI method, we run into several issues that are important and are the main results of this paper. These are:

1. **LoRaWAN Module**: LoRaWAN is designed for low-cost and, therefore, there is significant variability in the results using different modules. Each module has its peculiarities. Further work is required to make either a standard module for consistent results or a standard that when implemented by different manufactures results in similar results.
2. **RSSI Model Accuracy:** *meanRSSI* values fluctuate and depend upon the LoRaWAN module. As shown in Figure 6, our distance estimation model could be considered accurate except at distances between 300 and 400 m. Designing a better method or using a better module can resolve this problem.
3. **Battery Capacity:** Different battery capacitates to run the UAV's LoRaWAN module cause different *meanRSSI* values for short distances (below 300 m); hence, we recommend using the same battery capacity throughout the whole set of measurements.
4. **Antenna Direction:** The module antenna direction and position affect the *meanRSSI* values captured by the GSs. Thus, when building the distance model, the position and direction of the antenna need to be fixed for all UAVs during the distance estimation model and location estimation stage. Otherwise, $C$ and $L$ factors will change, resulting in inaccurate estimation of the distances, and thus, wrong locations.
5. **Seeeduino LoRaWAN Module Power Cable:** We found that the cable used to provide the power to the module attached to the UAV should be in the opposite direction of the antenna to balance the power in all directions. This issue is due to the fact that the cable can act as a second antenna for the Seeeduino LoRaWAN module, which affects the *meanRSSI* values for the modules in that direction of the UAV.
6. **Battery Location:** The battery used with the LoRaWAN module attached to the UAV needs to be under the module; otherwise, the *meanRSSI* values will be higher from the battery side, resulting in inaccurate distance estimation models.
7. **Environments:** Different environments affect the *meanRSSI* values because the model is based on a specific environment. This factor also results in different $L$ and $C$ values and thus, different distance estimation models. As a result, the values of $L$ and $C$ need to be calibrated to fix the difference in the *meanRSSI* value between the two environments.
8. **Modeling Range:** The *meanRSSI* values for shorter distances (less than 100 m) are not useable because of their high variability. If there are GSs located throughout a city, some GSs will be more than 100 m distance from a UAV; hence, it may not be a problem.
9. **Movement:** We had to keep the UAV stationary for at least ten seconds to get the *meanRSSI* to be used for distance estimation. This factor is because our LoRaWAN module required at least two seconds interval between successive messages, and we need five such messages to compute the *meanRSSI* value. A better module design may allow to overcome this and continuously measure the location.

10. **Underestimation**: In our experiments, we had the minimum number of GSs required for the 3-D location. In such cases, it is possible that the estimated distances are lower than actual, and the four spheres do not intersect. Mathematically, this shows up as a negative value under the square root resulting in "imaginary" height for the UAV.

## VI. Conclusions and Future Work

Remote IDs on UAVs will allow law-enforcement authorities to determine the ownership of the UAVs. Making the UAVs simply broadcasting their GPS-determined location may not be sufficient in all environments. In some situations, determining the location using the reception on ground stations is appropriate. In this paper, we proposed LoRaWAN as one possible wireless technology to use for Remote ID transmission and showed how ground stations could use *meanRSSI* values to determine the 3-D location of UAVs. We developed a prototype using commercially available low-cost LoRaWAN modules to identify and locate the UAVs, and we uncovered several issues that were documented in Section V. These are the main contributions of this paper. We plan to do further work to address these issues in the near future.

# Data Availability

The LoRaWAN modules configurations, Arduino IDE setup, and codes used to support the findings of this study are documented in [56].

# Conflicts of Interest

The authors declare that there is no conflict of interest regarding the publication of this paper.

## Funding Statement

Ali Ghubaish was funded by Prince Sattam Bin Abdulaziz University, AlKharj 11942, Saudi Arabia.

## Acknowledgments

I would like to thank my master thesis committee, Professor Roger Chamberlain and Ben Moseley, for their excellent comments and help to improve the overall write-up. In addition, I would like to thank my colleagues Guillaume Valentis, Xipeng Wang, Arghya Datta, Marcio Teixeira, Maede Zolanvari, Ria Das, and Yousef AlShehri for their help in this paper.